\documentclass[12pt]{iopart}

\usepackage{harvard}

\usepackage{graphicx}

\def\wig#1{\mathrel{\hbox{\hbox to 0pt{%
          \lower.6ex\hbox{$\sim$}\hss}\raise.4ex\hbox{$#1$}}}}

\def\mearth{\rm M_\oplus}

\def\mjup{$\rm M_J$}
\def\rjup{$\rm R_J$}

\def\object{}

\begin{document}

\title[The composition of exoplanets]{The composition of transiting giant extrasolar planets.}

\author{T Guillot}

\address{Laboratoire Cassiop\'ee, CNRS UMR 6202, Observatoire de la
  C\^ote d'Azur, BP4229, 06304 Nice Cedex 4} 

\ead{guillot@obs-nice.fr}

\date{Submitted to the {\it Proceedings of the 135th Nobel Symposium},
  \today} 

\begin{abstract}
In principle, the combined measurements of the mass and radius a giant
exoplanet allow one to determine the relative fraction of hydrogen and
helium and of heavy elements in the planet. However, uncertainties on
the underlying physics imply that some known transiting planets appear
anomalously large, and this generally prevent any firm conclusion when
a planet is considered on an individual basis. On the basis of a
sample of 9 transiting planets known at the time, Guillot et al. {\it
A\&A} {\bf 453}, L21 (1996), concluded that all planets could be
explained with the same set of hypotheses, either by large but
plausible modifications of the equations of state, opacities, or by
the addition of an energy source, probably related to the dissipation
of kinetic energy by tides. On this basis, they concluded that the
amount of heavy elements in close-in giant planets is correlated with
the metallicity of the parent star. Furthermore they showed that
planets around metal-rich stars can possess large amounts of heavy
elements, up to 100 Earth masses. These results are confirmed by
studying the present sample of 18 transiting planets with masses
between that of Saturn and twice the mass of Jupiter. 
\end{abstract}



\section{Introduction}

The discovery of exoplanets in transit in front of their parent star
has led to the birth of a new field in science: exoplanetology,
defined as the study of the characteristics, formation and fate of
planets outside our solar system. I will not venture into trying to
detail the possibility offered by these transiting systems, but they
are numerous. One will be of importance for us, the possibility to
measure relatively precisely both the mass and radius of these planets
and thus armed with theoretical models, to constrain their
compositions.

The first transiting exoplanet detected, HD209458b
\cite{Charbonneau+2000,Henry00}, was shown to have a radius of
1.35\,\rjup\ for a mass of 0.65\,\mjup\ ($\rm R_J =71492$km is
Jupiter's equatorial radius at 1 bar, and $\rm M_J$ is Jupiter's
mass). This large radius implied naturally that the planet was a gas
giant, mainly formed with hydrogen and helium
\cite{Guillot96,Burrows00}, and a sign that our understanding of
planet formation, badly shaken by the discovery of 51 Peg b, was not
completely off-track.

However, \object{HD209458b} was then shown to be anomalously large
\cite{Bodenheimer+2001,GuillotShowman2002}, i.e. larger than
predicted by the evolution of a solar composition gas with no addition
of any heavy elements. In contrast, when other transiting planets were
found, they appeared to be ``normal'' within the error bars
\cite{Baraffe+2005,Laughlin2005}, although some indications
that they could possess relatively large amounts of heavy elements
were found \cite{Guillot2005}. This is until
\object{HD149026b} was shown to be significantly smaller than
expected, requiring the presence of a large amount $\sim 70\mearth$ of
heavy elements in its interior \cite{Sato+2005,Fortney06}.

On the basis of the 9 transiting planets known at the time,
\citeasnoun{Guillot06} proposed to explain the structure of all planets
with the same hypotheses. This required two elements: invoking a
``missing physics'' (e.g. energy dissipation, a modification of
equations of state or opacities), with the same rules for every
planet, and allowing for a variable amount of heavy elements in their
interior. Their results indicated a likely correlation between the
mass of heavy elements in the planets and the metallicity of the
parent star. This was later confirmed by \citeasnoun{Burrows07}, on the
basis of a sample that had grown to 14 members, and using as ``missing
physics'' an increased opacity to slow the contraction of the
planets. 

This paper updates and extends the previous work by considering a
sample of 18 planets. I purposely leave out two transiting planets
with masses that are quite far from the Saturn-Jupiter mass range:
GJ436 \cite{Gillon07}, a $\sim 20\mearth$ planet, and HD147506 (aka
HAT-P-2) \cite{Bakos+2007b}, with its 8.2\,\mjup\ mass. Other planets
are shown in table~\ref{tab:transiting}. Section~\ref{sec:principle}
describes the principle of the calculations. I then show on the basis
of standard evolution calculations that anomalously large planets are
common, with strong implications for which mechanisms can be
responsible for the slowing of the
contraction. Section~\ref{sec:explaining} examines several hypotheses
for the ``missing physics'' and investigates the mass of heavy elements -
stellar metallicity correlation.

\begin{table*}[htbp]
\caption{Transiting planets included in this study}
\label{tab:transiting}
\centering
\small
  \begin{tabular}{lrrrrrr} \hline \hline
 Name & Age [Ga] & [Fe/H] & $T_{\rm eq,0}$ [K] & $M_{\rm p}$ $[\rm M_{J}]$ &  $R_{\rm p}$ $[\rm R_{J}]$ & Refs. \\
 {\bf HD209458}   & $4-7$     & $0.02(3)$   & $1469(120)$ & $0.69(20)$  & $1.320(25)$ & \tiny [Ch00]Wi05/Kn07 \\
 {\bf OGLE-TR-56} & $2-4$     & $0.25(8)$   & $2214(72)$  & $1.29(12)$  & $1.300(50)$ & \tiny [Ko03]To04/Po07 \\
 {\bf OGLE-TR-113}& $0.7-10$  & $0.15(10)$  & $1340(80)$  & $1.35(22)$  & $1.090(30)$ & \tiny [Bo04]Gi06 \\
 {\bf OGLE-TR-132}& $0.5-2$   & $0.37(7)$   & $2110(100)$ & $1.19(13)$  & $1.180(70)$ & \tiny [Bo04]Gi07  \\
 {\bf OGLE-TR-111}& $1.1-10$  & $0.19(7)$   & $1033(36)$  & $0.52(13)$  & $1.010(40)$ & \tiny [Po04]Sa06/Wi07/Mi07 \\
 {\bf OGLE-TR-10} & $1.1-5$   & $0.28(10)$  & $1509(80)$  & $0.61(13)$  & $1.122(100)$& \tiny [Ko05]Po07/Ho07 \\
 {\bf TrES-1}     & $2-6$     & $0.06(5)$   & $1156(140)$ & $0.75(7)$   & $1.081(29)$ & \tiny [Al04]So04/Wi07 \\
 {\bf HD149026}   & $1.2-2.8$ & $0.36(5)$   & $1740(150)$ & $0.33(3)$   & $0.726(64)$ & \tiny [Sa05]Ch06 \\
 {\bf HD189733}   & $0.5-10$  & $-0.03(4)$  & $1199(100)$ & $1.15(4)$   & $1.154(32)$ & \tiny [Bo05]Po07 \\
 {\bf XO-1}       & $0.65-8$  & $0.015(40)$ & $1255(100)$ & $0.90(7)$   & $1.184(25)$ & \tiny [Mc06]Ho06/M06\\
 {\bf HAT-P-1}    & $2.6-4.6$ & $0.13(2)$   & $1316(61)$  & $0.53(4)$   & $1.203(51)$ & \tiny [Ba07]Wi07 \\
 {\bf TrES-2}     & $2.8-7.8$ & $-0.15(10)$ & $1459(90)$  & $1.28(9)$   & $1.220(45)$ & \tiny [OD06]So07 \\
 {\bf WASP-1}     & $0.3-3$   & $0.26(3)$   & $1849(80)$  & $0.867(73)$ & $1.443(89)$ & \tiny [Ca06]Sh06/Ch06 \\
 {\bf WASP-2}     & $0.3-10$  & $?$         & $1293(136)$ & $0.88(7)$   & $1.038(50)$ & \tiny [Ca06]Ch06 \\
 {\bf XO-2}       & $1-3$     & $0.45(2)$   & $1316(21)$  & $0.57(6)$   & $0.973(30)$ & \tiny [Bu07] \\
 {\bf TrES-3}     & $0.3-10$  & $?$         & $1644(90)$  & $1.92(23)$  & $1.295(81)$ & \tiny [OD07] \\
 {\bf TrES-4}     & $0.3-10$  & $?$         & $1757(89)$  & $0.84(20)$  & $1.674(94)$ & \tiny [Ma07] \\
 {\bf HAT-P-3}    & $0.3-10$  & $0.27(4)$   & $1150(38)$  & $0.599(26)$ & $0.890(46)$ & \tiny [To07] \\
 {\bf HAT-P-4}    & $3.6-6.8$ & $0.24(8)$   & $1689(60)$  & $0.680(40)$ & $1.270(50)$ & \tiny [Ko07] \\
    \hline \hline
\multicolumn{7}{l}{\parbox{\textwidth}{The numbers in parenthesis represent the uncertainties on the
  corresponding last digits.}}\\
\multicolumn{7}{l}{\parbox{\textwidth}{$\rm M_{Jup}=1.8986112\times 10^{30}\,$g is the
  mass of Jupiter. $\rm R_{Jup}=71,492\,$km is Jupiter's equatorial radius.}}\smallskip\\
\multicolumn{7}{l}{\parbox{\textwidth}{References: \cite{Charbonneau+2000,Konacki2003,Bouchy2004,Pont2004,Torres2004,Alonso2004,Sozzetti+2004,Sato+2005,Bouchy2005,Winn2005,ODonovan+2006,CollierCameron+2006,Knutson+2007,Gillon2006,Charbonneau+2006,Holman2006,Shporer+2007,Winn+2007,Winn+2007b,Bakos+2007,Burke+2007,ODonovan+2007,Mandushev+2007,Torres+2007,Pont+2007,Gillon+2007,Minniti+2007,Winn+2007,Kovacs+2007?}; The discovery papers are in brackets. The table is taken from F. Pont's site: {\tt http://obswww.unige.ch/$\sim$pont/TRANSITS.htm}.}}
\normalsize
\end{tabular}
\end{table*}

\section{Principle and hypotheses of the calculations}
\label{sec:principle}

Once they have formed, giant planets contract monotonically and
quasi-statically. The model used to calculate their evolution is
described elsewhere \citeaffixed{Guillot2005}{see}. It assumes that
the planets are made of a central core which contributes relatively
little to the global cooling, and of an extended, solar-composition
envelope which accounts for most of the planetary size and internal
energy.

An important problem behind models of the physical evolution of giant
planets lies in the initial condition used and in their orbital
evolution. Fortunately, initially extended planets cool and contract
rapidly when the stellar irradiation is not dominant, so that initial
conditions are generally forgotten over timescales of $10^7$ to $10^8$
years \citeaffixed{Marley+2007}{see}. The situation may be more complex for
Pegasids because of their rather slow initial contraction
\cite{Guillot96}, and the possibility that they may have formed at
larger distances before being brought close to their stars
\cite{Burrows00}. 

To make things simple, I assume that all planets have an initial
radius of 2\,\rjup, have formed in situ, and underwent a constant
irradiation from their parent star. Although this hypothesis can
certainly be questioned, it has a limited impact on the results
\citeaffixed{Ikoma+2006}{e.g.}. Most importantly, applying the same set of
hypotheses consistently to all the transiting planets considered
will eventually enable testing them when a statistically sufficient
number of such planets will have been discovered.

The assumption of solar composition in the envelope may also be
surprising in regard to the fact that our giant planets appear to all
be significantly enriched (by factors of $\sim3$ for Jupiter to 30 for
Uranus and Neptune over the Solar value). At this point, this is a
clear simplification, to be further discussed in \S~\ref{sec:opacity}.

Important ingredients entering the calculations are the equations of
state. The well-known EOS by \citeasnoun{SCVH95} is used in the
envelope. This EOS includes only hydrogen and helium, but a slightly
larger mass fraction of helium ($Y=0.30$) is used to mimic the
presence of a solar composition of heavy elements. The dense core is
assumed to be made of pure rocks, and the EOS is based on the
experimental fit by \citeasnoun{HM89}. Unless otherwise stated, the mass of
the core is the only parameter to be modified in the calculations.

The opacities are based on a table calculated by
\citeasnoun{Allard+2001}. These opacities do not account for the possible
presence of clouds. The progressive growth of an external radiative
zone governs most of the evolution (in particular contraction) of
irradiated giant planets \cite{Guillot96}. I will come back on
how modifications to the opacities can change the analysis of the
compositions of Pegasids. 

\begin{figure}
  \centerline{\resizebox{14cm}{!}{\includegraphics{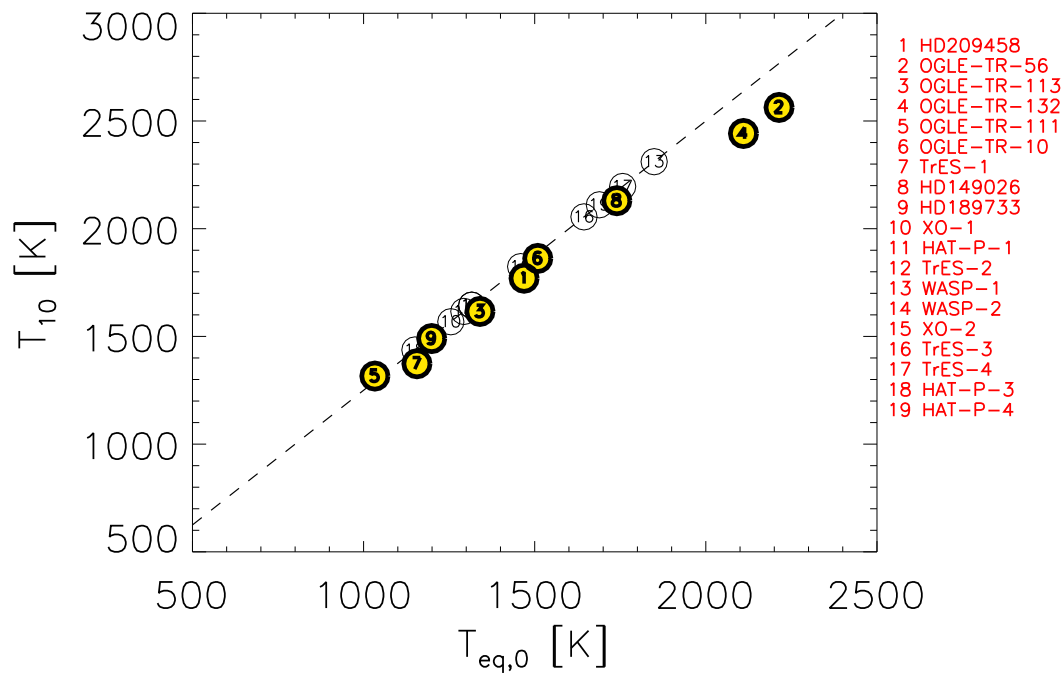}}}
  \caption{Temperature at the 10\,bar level as a function of the
  equilibrium temperature, as calculated with a 1D equilibrium model
  \cite{Iro+2005} (cases 1 to 9) and when assuming $T_{10}=1.25 T_{\rm
  eq,0}$ (cases 10 to 18).}
  \label{fig:teq}
\end{figure}

The atmospheric boundary condition used in these
calculations is simplified. In the case of the first 9 transiting
planets, I use specific atmospheric models calculated on the basis
of \citeasnoun{Iro+2005}. For the remaining planets, a simple relation is
used: 
\begin{equation}
T_{10}=1.25 T_{\rm eq,0},
\end{equation}
where $T_{10}$ is the temperature at the 10 bar level, and $T_{\rm
  eq,0}$ is the equilibrium temperature at the planet for a $A=0$
albedo (perfect absorption).  Figure~\ref{fig:teq} shows how this
approximate relation compares to the more elaborate but 1D radiative
transfer calculations. 

The difference between the observed transit radius (at $\sim $mbar
levels) and the modelled radius (at 10\,bar) has to be taken into
account \citeaffixed{Burrows+2003}{see}. As in \citeasnoun{Guillot06},
we include this radius difference as:
\begin{equation}
\Delta R =  {{\cal R} T_{\rm
    eq,0}\over \mu g} \ln \left({P_{\rm model}\over P_{\rm transit}}\right),
\label{eq:DR}
\end{equation}
where $P_{\rm model}=10$\,bar, we assume $P_{\rm transit}=1$\,mbar,
$\cal R$ is the gas constant, $\mu$ is the mean molecular weight (we
use $\mu=2.2$) and $g$ is the atmospheric gravity. This relation uses
a mean atmospheric temperature equal to $T_{\rm eq,0}$. Clearly
eq.~\ref{eq:DR} is an approximation, both from the point of view of
the transit pressure, and from that of the atmospheric temperature
structure. However, what is most important is that whatever the error
made here, it applies to {\it all planets in the same
way}. Furthermore, I claim that it is small compared to the inherent
uncertainties related to atmospheric models (including the possible
presence of clouds and their structure, the inhomogeneous heating and
the resulting dynamical effects).

\section{Anomalously large planets are common!}

\begin{figure}
  \centerline{\resizebox{14cm}{!}{\includegraphics{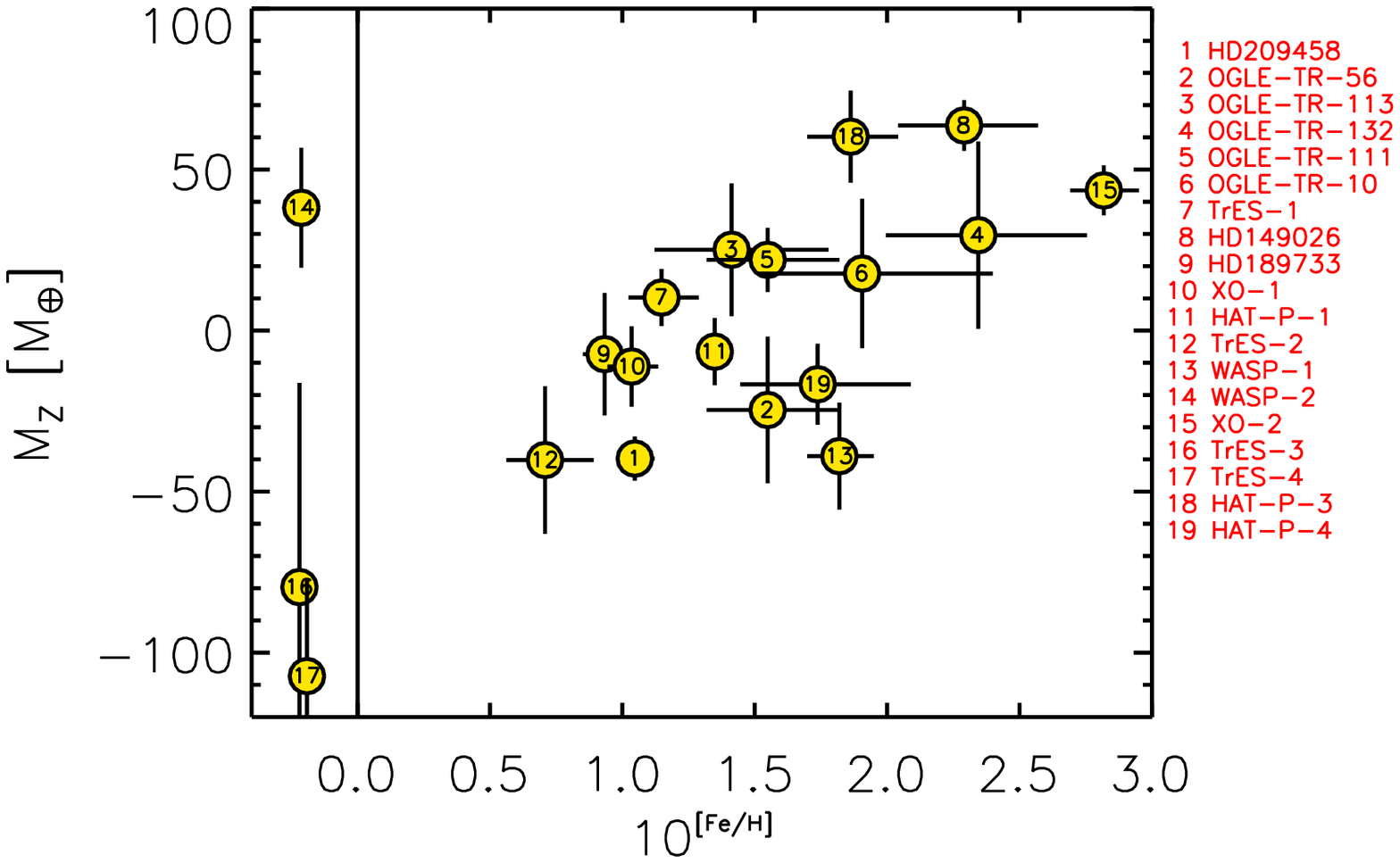}}}
  \caption{Mass of heavy elements in the planets as a function of the
    metal content of the parent star relative to the Sun. The mass of
    heavy elements required to fit the measured radii is calculated on
    the basis of standard evolution models. Negative core masses are
    required in some cases, implying that some significant physical
    input is missing (see text). Horizontal error bars correspond to the
    $1\sigma$ errors on the [Fe/H] determination. Vertical error bars
    are a consequence of the uncertainties on the measured planetary
    radii and ages.}
  \label{fig:std}
\end{figure}

Figure~\ref{fig:std} shows the masses of heavy elements that are
needed to explain the observed radius measurements, using a standard
evolution model. The results are plotted as a function of the heavy
element enrichment in the parent star (i.e. $10^{\rm [Fe/H]}$, where
[Fe/H] is the metallicity).

As already observed by many authors
\cite{Bodenheimer+2001,GuillotShowman2002,Chabrier+2004}, HD209458b
is anomalously large. In figure~\ref{fig:std}, this implies that the
planet appears with a {\it negative} core mass\footnote{Note that I
did not attempt to calculate hydrostatic models using negative core
masses: this is a result of the interpolation (in this case
extrapolation) technique used to obtain the core mass. As a result,
the precise value of $M_Z$ is of little significance when negative.},
which is of course physically impossible. Of interest to us is the
fact that a rather large number of planets (\object{HD209458b,
TrES-2b, WASP-1b, TrES-3b, TrES-4b, HAT-P-4b}, and probably also
\object{OGLE-TR-56b, HD189733b, XO-1b, HAT-P-1b}), 6 to 10 over 19,
appear in the negative core mass side of the diagram.
 
Clearly, this rules out mecanisms that are statistically unlikely as
responsible of the large radii. The small but non-zero eccentricity
explanation \cite{Bodenheimer+2001} is already excluded for HD209458b
\cite{Deming+2005,Laughlin2005b}. The spin-orbit trapping into a
Cassini state \cite{WinnHolman2005}, has been shown to be unprobable
\cite{Levrard+2007,Fabrycky+2007}.  On the other hand, the
semi-convection mechanism put forward by
\citeasnoun{ChabrierBaraffe2007} cannot be ruled out, but one would
expect it to work more efficiently at large metallicities, which seems
not to be compatible with the results of fig~\ref{fig:std}.

Altogether, this strengthens the case of a mechanism at work for {\it all}
planets, as discussed in previous papers
\cite{Guillot2005,Guillot06}. The rest of the article will be based on
this assumption: whatever happens to HD209458b or TrES-4b must also be
occuring for all other pegasids.

\section{Explaining the observations}
\label{sec:explaining}

\subsection{A hard equation of state model}

A first possibility is that the hydrogen helium equation of state
(EOS) used is not quite right and overestimates the density for a given
pressure, temperature and composition (we need hydrogen to be
``harder'', i.e. less compressible). Rather large uncertainties ($\sim
10-20\%$) of the density predicted by different equations of state are
possible at pressures and temperatures relevant to giant planet
interiors \cite{SaumonGuillot2004}. A $30\%$ {\it softening} of the
hydrogen EOS compared to the SCVH EOS has been obtained as a result of
ab-initio simulations matching high-pressure experiments (Militzer \&
Hubbard, personnal communication; see also
\citeasnoun{Militzer+2006}). Although this last result goes in the wrong
direction for us, it shows that the EOS is an important uncertainty
source.

\begin{figure}
\centerline{\resizebox{14cm}{!}{\includegraphics{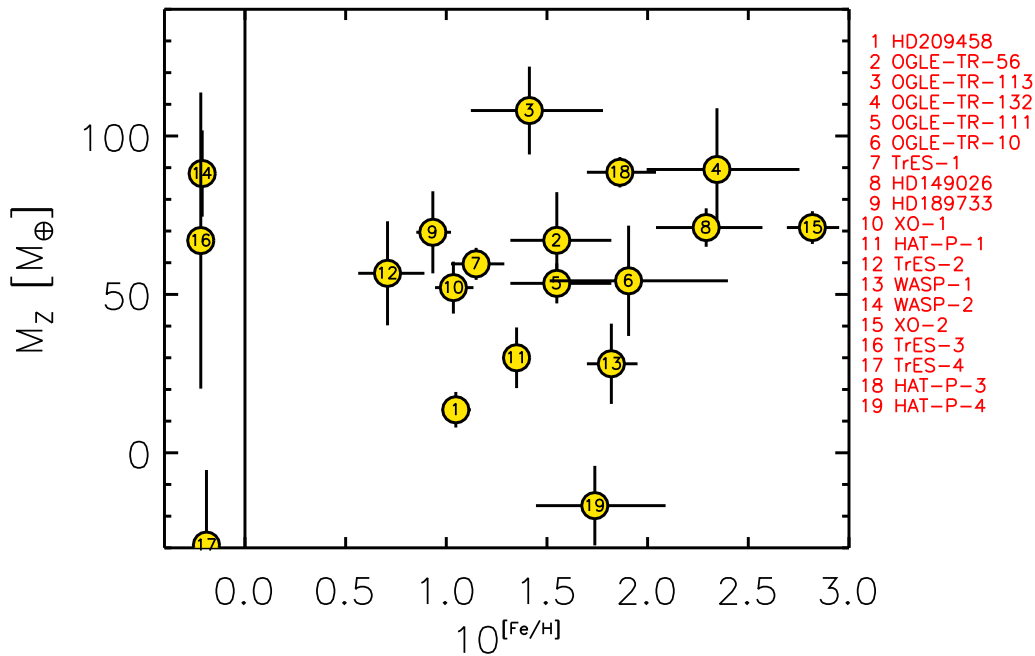}}}
\caption{Same as fig.~\ref{fig:std}, but the mass of heavy elements
  required to fit the measured radii are calculated on the basis of
  evolution models of a planet with {\it} no helium, to mimic a hard
  hydrogen EOS (see text).}
\label{fig:y0}
\end{figure}

In order to test the possibility that the EOS may be the culprit in
the underestimations of theoretical planetary radii, I present in
fig.~\ref{fig:y0} models calculated with the SCVH EOS, but with helium
removed. This is similar in essence to a $\sim 30\%$ hardening of the
EOS. 

As a result, we find that this large change to the EOS is sufficient
to explain the present radii of all planets except TrES-4b and
HAT-P-4b. However, these two planets have been discovered recently,
and still have relatively large error bars. Future measurements are
needed for these objects. 

Globally, the trend between $M_Z$ and [Fe/H] that was seen in the
standard model is mostly lost: this effect is an artificial one that
is related to the different adiabats of hydrogen and helium. As a
result, the change in radius between a pure H planet and a H/He
mixture is greater for planets that receive less flux from their
star. It shows however that both changes in the density and specific
entropies have to be carefully accounted for in new EOSs.

\subsection{An increased opacity scenario}
\label{sec:opacity}

Another possibility to explain the anomalously large planets is a
serious underestimation of the Rosseland opacities. The growth of the
external radiative region (down to $\sim$kbar pressure levels)
controls the cooling and hence the contraction of all pegasids
\cite{Guillot96,GuillotShowman2002}. This region is in a difficult,
relatively cold but dense region, and generally, opacities are
notoriously famous for the large uncertainties associated to them. 

\begin{figure}
\centerline{\resizebox{14cm}{!}{\includegraphics{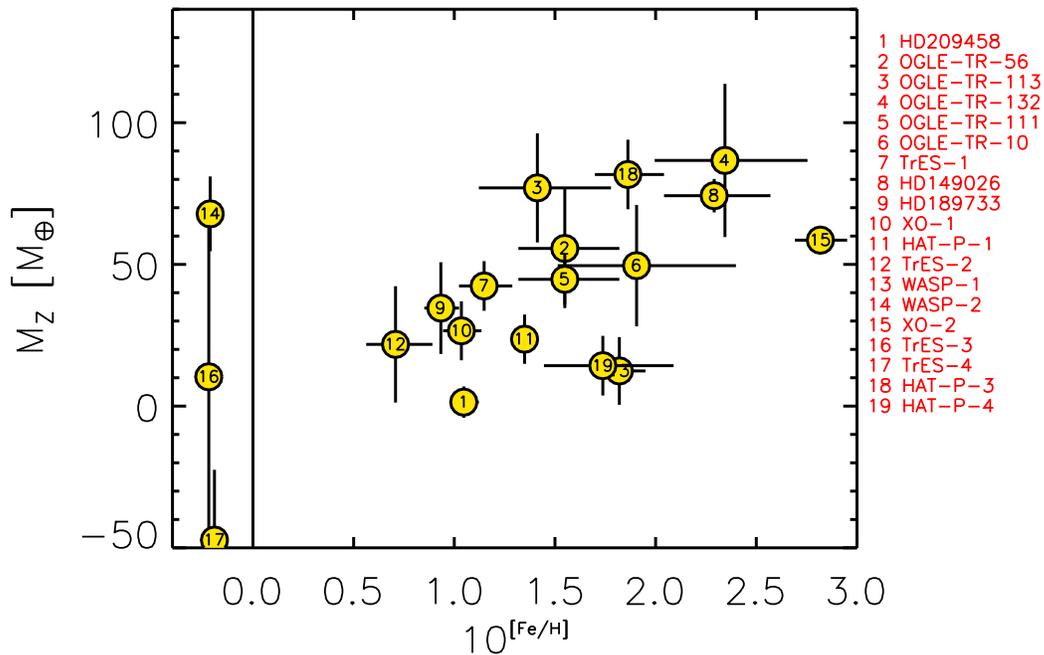}}}
\caption{Same as fig.~\ref{fig:std}, but the mass of heavy elements
  required to fit the measured radii are calculated on the basis of
  evolution models with opacities arbitrarily increased by a factor
  $30$.}
\label{fig:opa30}
\end{figure}

Figure~\ref{fig:opa30} presents results obtained with Rosseland
opacities that have been arbitrarily multiplied by a factor 30
compared to the fiducial values. This value corresponds
approximatively to the increase that is needed to explain the radius
measured for HD209458b. 

It can be seen that this approach leads to physically sound values of
the heavy element masses for all planets, with two exceptions: TrES-4b
and HAT-P-3b. It may be premature however to rule out this explanation
because these discoveries are quite recent, and experience has shown
that the inferred radii which are very dependent on stellar properties
have often moved a bit more than would be expected from their
$1-\sigma$ error bar. It is clear that TrES-4b and HAT-P-3b are
extremely interesting planets to test these models. More discoveries
will of course also help to settle the matter.

Interestingly, we again observe a clear correlation between planetary
and stellar ``metallicity''. This result, first found by
\citeasnoun{Guillot06}, was later confirmed by \citeasnoun{Burrows07}.
A preliminary version of the last paper, put onto astro-ph,
claimed that this opacity increase was related to an oversolar
enrichement of the envelope, a simple, logical explanation when
considering that the atmospheres of our giant planets are enriched in
heavy elements. However this explanation was based on a model for
which only opacities had been increased, and not the mean molecular
weight. A consistent treatment, unfortunately, rules out this simple
explanation as a solution to the anomalously large planets problem. 

\begin{figure}[htbp]
\centerline{\resizebox{14cm}{!}{\includegraphics[angle=0]{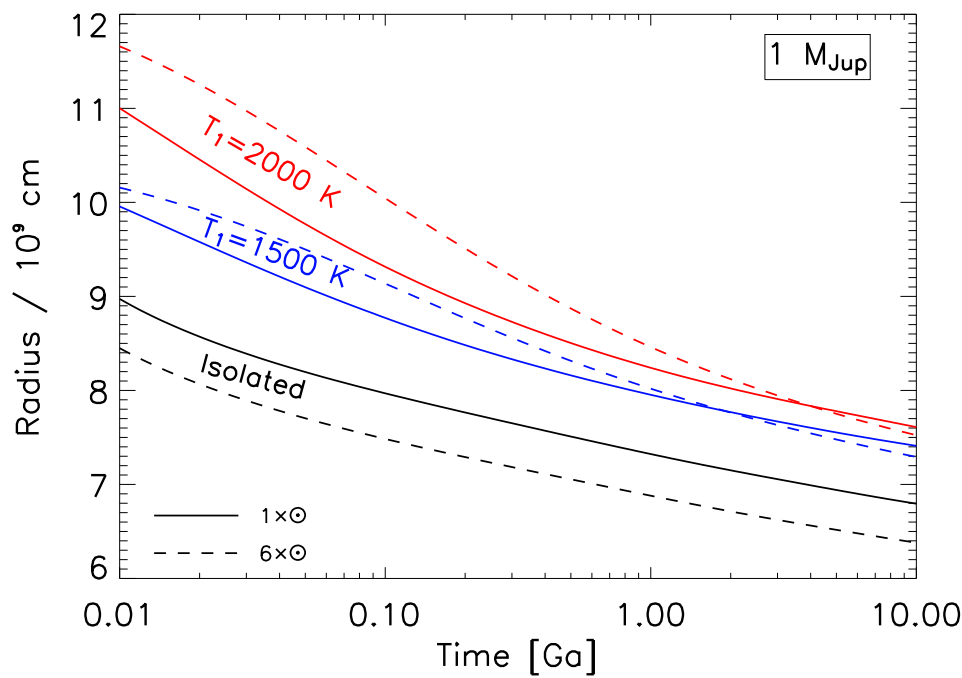}}}
\caption{Evolution of giant planets in terms of radius vs. time, for
  different irradiation levels, and 2 assumed compositions: solar, and
  6 times solar. (This calculation ignores second order effects as
  modifications of the adiabatic temperature gradient and
  non-linear effects in the opacity calculation, and more importantly
  modifications of atmospheric properties.) [From \citeasnoun{Guillot2005}].}
\label{fig:evols} 
\end{figure}

The reason for that is a balancing effect between the slower cooling
and the increased mean molecular weight that implies that enriching
the envelopes in heavy elements does not generally make the planets
larger, and often results in the oposite
effect. Figure~\ref{fig:evols} shows that the mean molecular weight
effect eventually wins, and that this occurs over a timescale similar
to the ages of the systems that are observed. 

Could the atmosphere or the upper part of the envelope be enriched in
heavy elements while the rest would not? This is regarded as extremely
unlikely, because a double-diffusive instability (``salt-fingers'')
should set in \citeaffixed{Stevenson1985}{e.g.}.

Finally, there is an interesting possibility that the envelopes of
giant planets orbiting metal-rich stars are more enriched than for
planets orbiting metal-poor stars. This case would lead to a {\it
stronger} correlation between heavy elements in the planet and stellar
metallicity.

\subsection{The kinetic energy mechanism}

The third mechansim that would be in action for all the planets, but
at a level that depends on the irradiation that they receive from the
parent star is the one proposed by \citeasnoun{ShowmanGuillot2002}: a
fraction ($\sim 1\%$) of the energy recieved in form of stellar
irradiation by the planet is converted into kinetic energy in the
atmosphere. This energy is transported to deeper levels, and
dissipated probably due to tidal interactions with the star. Although
simulations and observations on Earth have shown this amount of
kinetic energy to be generated in the atmosphere, several problems
remain with this scenario: first the turbulent viscosity needs to be
small otherwise dissipation takes place in the upper atmosphere, with
little effect on the planet's contraction \cite{Burkert05}; Second
kinetic energy must be transported to the 10-100bar level at least;
Third, it must be dissipated there (or at deeper levels).

\begin{figure}
\centerline{\resizebox{14cm}{!}{\includegraphics{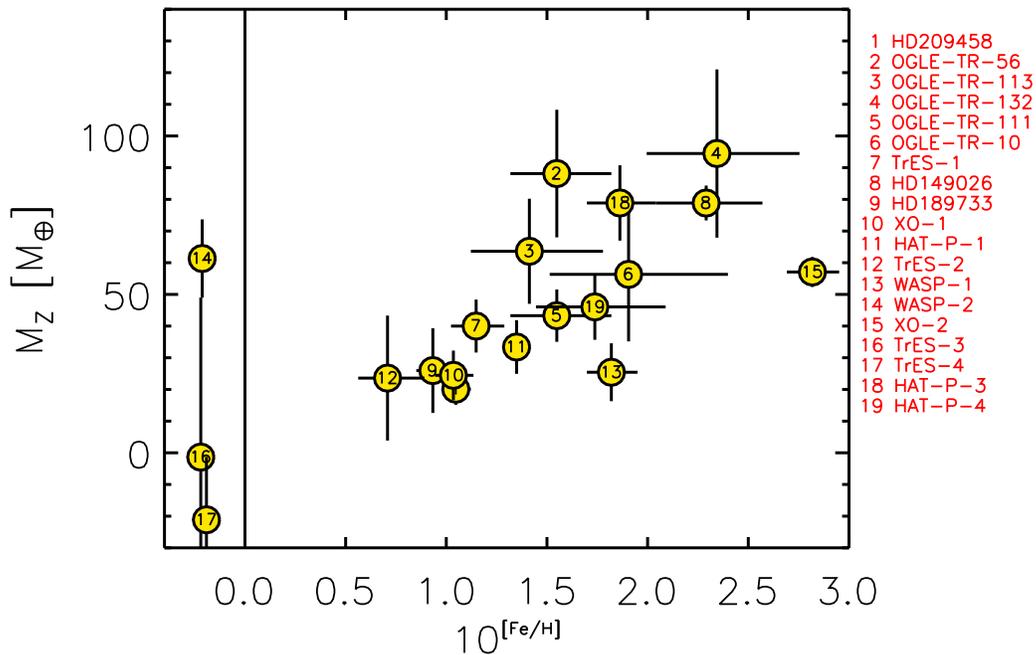}}}
\caption{Same as fig.~\ref{fig:std}, but the mass of heavy elements
  required to fit the measured radii are calculated on the basis of
  evolution models including an additional heat source slowing the
  cooling of the planet. This heat source is assumed equal to $0.5\%$
  of the incoming stellar heat flux \cite{ShowmanGuillot2002}.}
\label{fig:diss}
\end{figure}

Figure~\ref{fig:diss} shows results obtained with this model, assuming
a $0.5\%$ dissipation of the incoming stellar flux at the center of
the planet. This level of dissipation is chosen so as to yield a small
but positive value for the mass of heavy elements inside
HD209458b. (Larger values lead to systematically larger masses of
heavy elements derived in all transiting planets.) As previously
mentioned \cite{Guillot06}, the relation between the mass of heavy
elements in the planet and the stellar metallicity is qualitatively
similar than when increasing opacities. 

\section{Conclusions}
\label{conclusions}

With the 18 transiting exoplanets with masses between that of Saturn
and twice the mass of Jupiter known thus far, we are able to confirm
the conclusions derived by \citeasnoun{Guillot06}:
\begin{itemize}
\item All planets can be explained in the framework of a single
  scenario, when accounting for some missing physics which can either
  be due to modifications in the EOSs, opacities, or to an additional
  heating source. 
\item Giant planets can possess a large mass in heavy elements, up to
  100 Earth masses (a value which is itself model-dependent). 
\item There is a correlation between the mass of heavy elements in the
  planets and the metallicity of their parent stars, so that the most
  massive planets generally orbit the most massive stars. This
  correlation is probably {\it not} a single relation, but shows, as
  would be expected, some scatter. However, there is presently an {\it
  absence} of planets with small masses of heavy elements around
  metal-rich stars, planets that appear to be predicted by formation
  models \citeaffixed{Alibert+2005}{see}.
\item Planets around stars that have metallicities comparable to that
  of our Sun generally tend to have very small cores. There is a clear
  lack of giant planets at short periods ($P<10$ days) around
  metal-poor stars \cite{Fressin07}. This seems to imply that
  metal-poor stars are not able to form giant planets and have them
  migrate very close-in.
\end{itemize}

It should be noted however that a planet such as TrES-4 may pose a
problem for these models if its large radius is confirmed. It is
presently too large to be explained by the models, but the
error bars are currently quite large and prevent any definitive
conclusion to be made. Future detections of both large and small
transiting pegasids are extremely important to test the models and
better understand how giant planets formed.

\ack
This research was supported by the {\it Programme National de
  Plan\'etologie}. The author thanks an anonymous referee for helpful
comments that have improved the manuscript.

\section*{References}
\bibliographystyle{jphysicsB}
\bibliography{stockholm07_guillot}

\end{document}